\documentclass[aps,prd,twocolumn,superscriptaddress,nofootinbib]{revtex4-2}

\usepackage{latexsym}
\usepackage{amsmath}
\usepackage{amssymb}
\usepackage{amsfonts}
\usepackage{bm}
\RequirePackage{lineno}

\usepackage{color}
\definecolor{purple}{rgb}{0.5,0,0.5}
\definecolor{blue}{rgb}{0.0,0,0.9}
\definecolor{prdblue}{rgb}{0.133,0.118,0.498}
\usepackage[colorlinks=true, pdfstartview=FitV, linkcolor=prdblue, citecolor= prdblue, urlcolor=prdblue]{hyperref}

\usepackage{supertabular} 
\usepackage{placeins}
\usepackage{epsfig}
\usepackage{graphicx}
\usepackage{hyperref}

\hypersetup{
  breaklinks=true,
  colorlinks = true,
  linkcolor = blue,
  anchorcolor = blue,
  citecolor = blue,
  filecolor = blue,
  pagecolor = blue,
  urlcolor = blue
}
\hypersetup{
  bookmarks=true,
  bookmarksnumbered=true,
  bookmarkstype=toc,
  linktocpage=true
}

\newcommand{\BESIIIorcid}[1]{\href{https://orcid.org/#1}{\hspace*{0.1em}\raisebox{-0.45ex}{\includegraphics[width=1em]{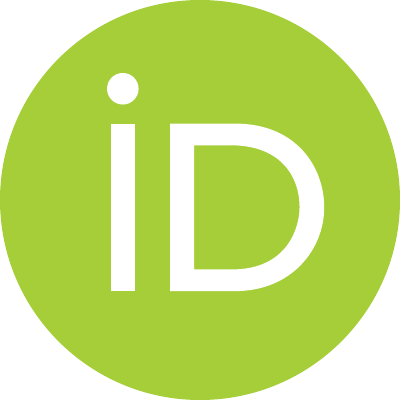}}}}
\begin{document}

\modulolinenumbers[2]

\setlength{\oddsidemargin}{-0.5cm} \addtolength{\topmargin}{15mm}

\title{\boldmath First Measurement of the $D_s^+\rightarrow K^{*}(892)^0\mu^+\nu_{\mu}$ Decay, Study of Dynamics and Test of Lepton Universality with $D_s^+\rightarrow K^{*}(892)^0\ell^+\nu_{\ell}$ Decays}

\author{
M.~Ablikim$^{1}$\BESIIIorcid{0000-0002-3935-619X},
M.~N.~Achasov$^{4,b}$\BESIIIorcid{0000-0002-9400-8622},
P.~Adlarson$^{81}$\BESIIIorcid{0000-0001-6280-3851},
X.~C.~Ai$^{87}$\BESIIIorcid{0000-0003-3856-2415},
C.~S.~Akondi$^{31A,31B}$\BESIIIorcid{0000-0001-6303-5217},
R.~Aliberti$^{39}$\BESIIIorcid{0000-0003-3500-4012},
A.~Amoroso$^{80A,80C}$\BESIIIorcid{0000-0002-3095-8610},
Q.~An$^{77,64,\dagger}$,
Y.~H.~An$^{87}$\BESIIIorcid{0009-0008-3419-0849},
Y.~Bai$^{62}$\BESIIIorcid{0000-0001-6593-5665}
% ...
 \\
 \vspace{0.2cm}
 (BESIII Collaboration)\\
 \vspace{0.2cm} {\it
$^{1}$ Institute of High Energy Physics, Beijing 100049, People's Republic of China\\
% ...
$^{4}$ Budker Institute of Nuclear Physics SB RAS (BINP), Novosibirsk 630090, Russia\\
% ...
$^{31}$ INFN Sezione di Ferrara, (A)INFN Sezione di Ferrara, I-44122, Ferrara, Italy; (B)University of Ferrara, I-44122, Ferrara, Italy\\
% ...
$^{39}$ Johannes Gutenberg University of Mainz, Johann-Joachim-Becher-Weg 45, D-55099 Mainz, Germany\\
% ...
$^{62}$ Southeast University, Nanjing 211100, People's Republic of China\\
% ...
$^{64}$ State Key Laboratory of Particle Detection and Electronics, Beijing 100049, Hefei 230026, People's Republic of China\\
% ...
$^{77}$ University of Science and Technology of China, Hefei 230026, People's Republic of China\\
% ...
$^{80}$ University of Turin and INFN, (A)University of Turin, I-10125, Turin, Italy; (B)University of Eastern Piedmont, I-15121, Alessandria, Italy; (C)INFN, I-10125, Turin, Italy\\
$^{81}$ Uppsala University, Box 516, SE-75120 Uppsala, Sweden\\
% ...
$^{87}$ Zhengzhou University, Zhengzhou 450001, People's Republic of China\\
\vspace{0.2cm}
$^{\dagger}$ Deceased\\
% ...
$^{b}$ Also at the Novosibirsk State University, Novosibirsk, 630090, Russia\\
% ...
\vspace{0.4cm}
}
}
\hspace{0.2cm}
\begin{abstract}
We report the first measurement of the semileptonic decay $D^+_s \rightarrow K^*(892)^0\mu^+\nu_{\mu}$ and an improved measurement of the decay $D^+_s \rightarrow K^*(892)^0 e^+\nu_{e}$ using a sample of $7.33~\mathrm{fb}^{-1}$ of $e^+e^-$ annihilation data collected at center-of-mass energies between 4.128 to 4.226~GeV with the BESIII detector at the BEPCII collider. We measure the branching fractions to be $\mathcal B({D^+_s\rightarrow K^*(892)^0 \mu^+\nu_{\mu}})=(2.07\pm0.22_{\rm stat}\pm0.10_{\rm syst})\times10^{-3}$ and $\mathcal B({D^+_s\rightarrow K^*(892)^0 e^+\nu_{e}})=(2.14\pm0.18_{\rm stat}\pm0.10_{\rm syst})\times10^{-3}$. Based on a simultaneous study of the dynamics in two semileptonic decays, the hadronic form factor parameters in the $D^+_s\rightarrow K^{*}(892)^0$ transition are determined to be $r_{V} = V(0)/A_1(0) = 1.63 \pm 0.14_{\rm stat} \pm 0.08_{\rm syst}$, $r_{2} = A_2(0)/A_1(0) = 0.60 \pm 0.13_{\rm stat} \pm 0.06_{\rm syst}$, and $A_1(0)=0.56 \pm 0.02_{\rm stat} \pm 0.01_{\rm syst}$, where $V(0)$ is the vector form factor and $A_{1,2}(0)$ are the axial-vector form factors evaluated at $q^2=0$. The precision of $r_V$ and $r_2$ is improved by twofold and $A_1(0)$ is measured for the first time. We also report the first model-independent measurements of the differential decay rates and the lepton forward-backward asymmetries for $D^+_s\rightarrow K^{*}(892)^0\ell^+\nu_{\ell}$ decays. Based on these measurements, we perform a test of lepton flavor universality in full and separate $q^2$ intervals with $D^+_s\rightarrow K^{*}(892)^0\ell^+\nu_{\ell}$ decays. No violation is found within uncertainties. Our results present for the first time a complete study of the dynamics in the $D_s^+\rightarrow K^*(892)^0$ transition, and provide stringent tests of various non-perturbative theoretical calculations.
\end{abstract}


\maketitle

Semileptonic (SL) decays of $D_s^+$ mesons provide valuable information on the weak and strong interactions of hadrons composed of heavy quarks~\cite{physrept494,RevModPhys67_893}. Their partial decay rate is related to the product of a hadronic form factor (FF) describing the strong interaction in the initial and final hadrons, and the Cabibbo-Kobayashi-Maskawa (CKM) matrix~\cite{prl10_531} element $|V_{cs(d)}|$ parametrizing the mixing between different flavors of quarks in the weak interaction. Since $|V_{cs(d)}|$ is tightly constrained by CKM unitarity, studies of $D_s^+$ meson SL decays provide an ideal opportunity to determine the hadronic FFs. 
In recent years, there has been a great deal of attention on experimental studies of hadronic FFs in $D_s^+\rightarrow V\ell^+\nu_{\ell}$ decays~\cite{pdg24}, where $V$ and $\ell$ refer to a vector meson and a lepton, respectively. However, for the $D_s^+\rightarrow K^{*}(892)^0$ transition, only $D_s^+\rightarrow K^{*}(892)^0e^+\nu_e$ has been experimentally reported~\cite{prl122_061801}.
The SL decay $D_s^+\rightarrow K^{*}(892)^0\mu^+\nu_{\mu}$ plays an important role in understanding the $D_s^+\rightarrow K^{*}(892)^0$ transition.

Theoretical calculations of FFs in $D^+_s\rightarrow K^{*}(892)^0$ transition are extensively carried out with a variety of non-perturbative approaches~\cite{PRD62_014006,PRD72_034029,IJMPA21_6125,PRD111_113005,PRD111_113005,JPG39_025005,EPJC77_587,PRD89_034013,PRD92_054038,PRD98_114031,FrontPhys14_64401,PRD101_013004,PRD109_026008,JHEP02_179,PRD78_054002}, which are important for understanding the quantum-chromodynamics (QCD) theory in the charm sector.
In these calculations, one vector FF, $V(q^2)$, and two axial-vector FFs, $A_{1,2}(q^2)$, are introduced to describe strong interaction effects. 
The FF parameters $r_V=V(0)/A_1(0)$, $r_2=A_2(0)/A_1(0)$, and $A_1(0)$ in the $D_s^+\rightarrow K^{*}(892)^0$ transition predicted by these  models~\cite{PRD62_014006,PRD72_034029,IJMPA21_6125,PRD111_113005,PRD111_113005,JPG39_025005,EPJC77_587,PRD89_034013,PRD92_054038,PRD98_114031,FrontPhys14_64401,PRD101_013004,PRD109_026008,JHEP02_179,PRD78_054002} differ significantly, varying from $1.31-1.93$, $0.62-0.99$, and $0.42-0.61$, respectively. 
In particular, the recent calculation for $A_1(0)$ within the framework of the light-cone sum rules (LCSR)~\cite{PRD111_113005} incorporating higher-twist corrections shows a significantly lower value compared to the previous LCSR calculation in Ref.~\cite{IJMPA21_6125}. 
Currently, the experimental precision on $r_V$ and $r_2$ is still poor~\cite{pdg24}, and no direct measurement of $A_1(0)$ is available for $D^+_s\rightarrow K^{*}(892)^0\ell^+\nu_{\ell}$ decays~\cite{pdg24}. Therefore, precise measurements of these FF parameters can provide stringent tests and calibrations of the various non-perturbative calculations, including Lattice QCD (LQCD).
Furthermore, LQCD predicts that FFs have minimal dependence on the spectator-quark mass, with FF values in
$D^+_s\rightarrow K^*(892)^0$ and $D^{+,0} \rightarrow \rho^{0,-}$ transitions differing by less than 5\%~\cite{lattice}.  
Verifying this prediction at the few percent level in charm SL decays will add confidence when applying LQCD to $B$ SL decays for precise determination of the CKM matrix elements $|V_{cb}|$ and $|V_{ub}|$~\cite{lattice,EPJC74_2981,prd85_114502}.

In the Standard Model, SL decays of $D_s^+$ mesons offer an excellent opportunity to test lepton flavor universality (LFU)~\cite{ARNPS,NSR,PRD91_094009,CPC45_063107,PRD96_016017,PRR2,IJMA40_2550118}. 
It has been suggested that observable LFU violation effects may occur in charm hadron SL decays~\cite{CPC45_063107,PRD91_094009,PRD96_016017,IJMA40_2550118,EPJC80_153}. 
In recent years, a range of theoretical models calculated the branching fractions (BFs) of the $D^+_s\rightarrow K^*(892)^0$ transition for both $\mu$ and $e$ modes. 
The predicted BFs vary by $(0.15-0.23)\%$~\cite{PRD62_014006,PRD72_034029,IJMPA21_6125,PRD111_113005,PRD111_113005,JPG39_025005,EPJC77_587,PRD89_034013,PRD92_054038,PRD98_114031,FrontPhys14_64401,PRD101_013004,PRD109_026008,JHEP02_179,PRD78_054002}, and the ratio of their BFs, $R^{\mu/e}_{K^*(892)}=\frac{\mathcal{B}(D^+_s\rightarrow K^*(892)^0\mu^+\nu_{\mu})}{\mathcal{B}(D^+_s\rightarrow K^*(892)^0e^+\nu_{e})}$ ranges from $0.93-0.99$~\cite{PRD62_014006,PRD72_034029,IJMPA21_6125,PRD111_113005,PRD111_113005,JPG39_025005,EPJC77_587,PRD89_034013,PRD92_054038,PRD98_114031,FrontPhys14_64401,PRD101_013004,PRD109_026008,JHEP02_179,PRD78_054002}.  
Measuring the ratio of their differential decay rates in the full range and separate kinematic regions provides an ideal test of LFU.
In particular, the presence of both vector and axial-vector couplings of electroweak bosons to fermions in the $D^+_s\rightarrow K^*(892)^0$ transition gives rise to an asymmetry in the polar angle of the lepton momentum. This forward-backward asymmetry, $A_{\rm FB}$, which is expected to be insensitive to FF uncertainties, can provide the other key measurement in a test of $\mu-e$ LFU. Therefore, study of the dynamics in $D^+_s\rightarrow K^{*}(892)^0\ell^+\nu_{\ell}$ decays provide critical data to distinguish among various theories and also test $\mu-e$ LFU. Furthermore, the multiple polarization states of vector meson in $D^+_s\rightarrow K^{*}(892)^0$ transition provide a more detailed search for potential new physics effects. The ratio  $\Gamma_L/\Gamma_T$ of the decay rates between the longitudinally and transversely polarized $K^*(892)^0$ fractions is sensitive to the pseudoscalar Wilson coefficient $c_P^{(\ell)}$~\cite{PRD91_094009}.  No such measurement has been reported yet~\cite{pdg24}. 

In this Letter, we report the first measurement of the absolute BF for $D^+_s\rightarrow K^*(892)^0\mu^+\nu_{\mu}$ and an improved measurement of the BF for $D^+_s\rightarrow K^*(892)^0 e^+\nu_{e}$, as well as the most precise determinations of the FF parameters $r_V$, $r_2$, and $A_1(0)$ in $D^+_s\rightarrow K^*(892)^0\ell^+\nu_{\ell}$ decays.
We also report the first model-independent measurements of the differential decay rates and the forward-backward asymmetries in $D^+_s\rightarrow K^*(892)^0$ transitions in the full kinematic range as well as individual four-momentum transfer regions. Using these observables, we provide for the first time a test of LFU using $D^+_s\rightarrow K^*(892)^0\ell^+\nu_{\ell}$ decays. 
Throughout this paper, charge-conjugate modes are implied unless explicitly noted.
These measurements are performed using an $e^+e^-$ annihilation data sample corresponding to an integrated luminosity of $7.33~\mathrm{fb}^{-1}$ produced at the center-of-mass energies between $\sqrt{s}=4.128$ and 4.226~GeV with the BEPCII collider and collected by the BESIII detector~\cite{Ablikim:2009aa,Yu:IPAC2016-TUYA01}.
The cross section for the production of $e^+e^-\rightarrow D_s^{*+}D_s^{-}$ is about 1~nb at energies around $\sqrt{s}=4.178$~GeV~\cite{prd80_072001}.

Monte Carlo (MC) simulated data samples produced with a {\sc geant4}-based~\cite{geant4} software package,
which includes the geometric description of the BESIII detector and the detector response,
are used to determine detection efficiencies and to estimate background contributions.
The inclusive MC sample includes the production of open charm processes, the initial state radiation (ISR) 
production of vector charmonium(-like) states, and the continuum processes incorporated in {\sc kkmc}~\cite{kkmc}. 
All particle decays are modeled with {\sc evtgen}~\cite{nima462_152} using branching fractions either taken from
the Particle Data Group~\cite{pdg24}, when available, or otherwise estimated with {\sc lundcharm}~\cite{lundcharm}.
Final state radiation (FSR) from charged final state particles is incorporated using the
{\sc photos} package~\cite{plb303_163}. The generation of signal $D^+_s\rightarrow K^*(892)^0\ell^+\nu_{\ell}$ incorporates knowledge of the form factors obtained in this work.

\begin{figure}[tp!]
\flushleft
\begin{center}
   \includegraphics[width=\linewidth]{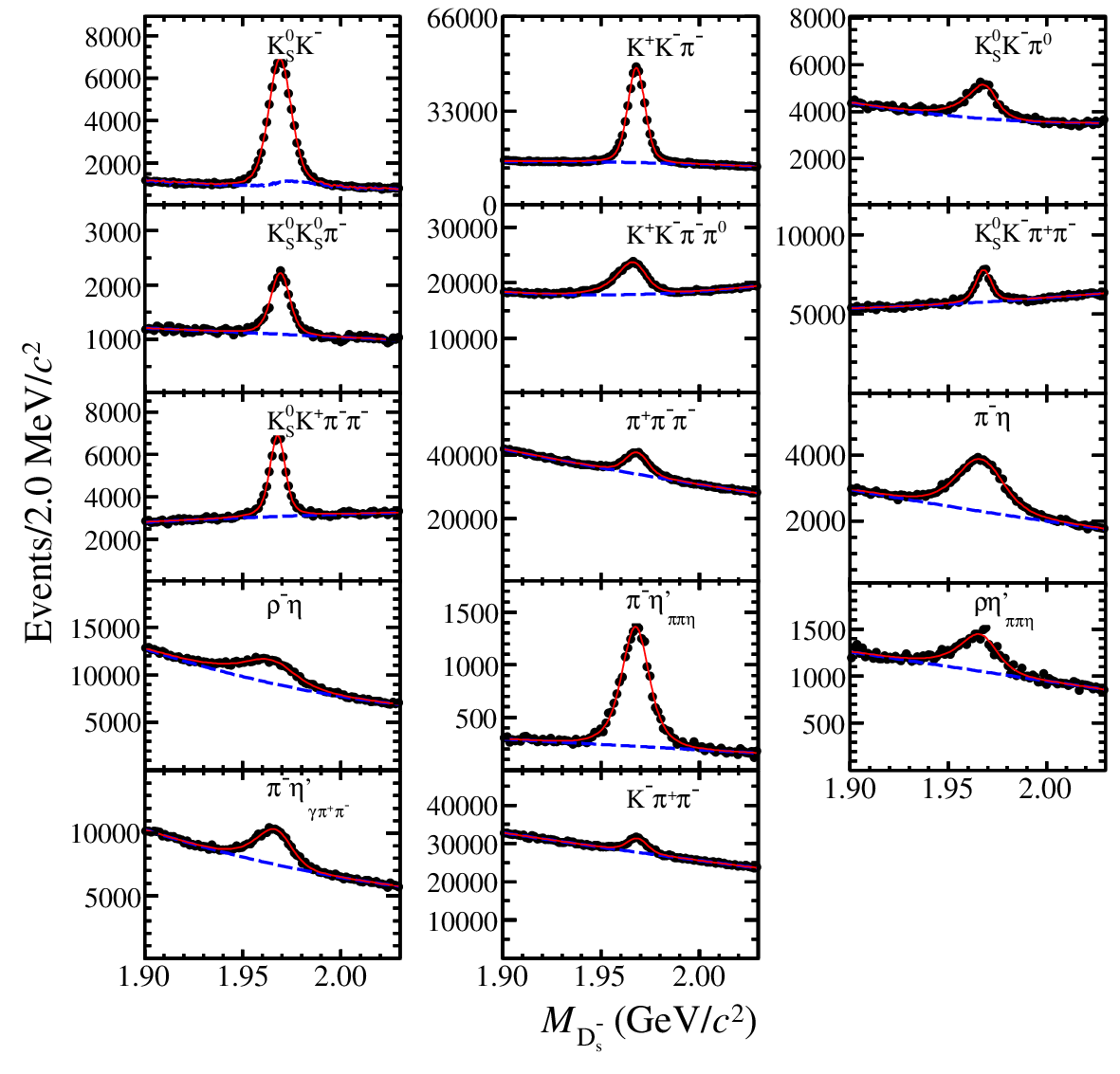}
\caption{(Color online)
Fits to $M_{D_s^-}$ distributions for the fourteen tag modes. The dots with error bars are data, dashed blue curves are the fitted background events and 
solid red curves are the total fits.}
\label{fig:tag_mds}
\end{center}
\end{figure}

This analysis uses both ``single-tag'' (ST) and ``double-tag'' (DT) samples of $D_s$ decays~\cite{prl122_061801}.
The STs are $D_s^-$ mesons reconstructed from their daughter particles in one of fourteen hadronic decays, as shown in Fig.~\ref{fig:tag_mds}, while the DTs are events with a ST and a reconstructed $D^+_s\rightarrow K^*(892)^0\ell^+\nu_{\ell}$ candidate. 
The BF for the SL decay is given by
\begin{equation}
  \mathcal{B}_{\rm SL} \,=\,
  \frac{N_{\rm DT}}{\sum_i N^{i}_{\rm ST} \,
      \left(\epsilon^i_{\rm DT}/\epsilon^i_{\rm ST}\right)} \,=\,
  \frac{N_{\rm DT}}{N_{\rm ST} \, \epsilon_{\rm SL}}, \label{eq:branch}
\end{equation}
where $N_{\rm DT}$ is the total yield of DT events, $N_{\rm ST}$ is the total ST yield, and 
$\epsilon_{\rm SL}=(\sum_i N^{i}_{\rm ST} \, \epsilon^i_{\rm DT}/\epsilon^i_{\rm ST})/\sum_i N^{i}_{\rm ST}$
is the average efficiency for reconstructing the SL decay in a ST event,
weighted by the measured yields of tag modes in the data, where $\epsilon^i_{\rm ST}$ and $\epsilon^i_{\rm DT}$ are the efficiencies for finding the ST and the SL decay in the $i$-th tag mode, respectively.

A detailed description of the selection criteria for $\pi^{\pm}$, $K^{\pm}$ and $\gamma$ candidates is given in Ref.~\cite{D0Kspiev}. 
The $\pi^0$, $\eta$, $K^0_S$, $\rho^-$, $\eta$ and $\eta^{\prime}$ mesons are reconstructed with the same technique as in Ref.~\cite{prd110_052012}.
In the case of multiple candidates in ST reconstruction, only the candidate with the recoil mass, $M_{\rm rec}$, closest to the $D_s^{*+}$ known mass~\cite{pdg24} is retained. The recoil mass is calculated as:
\begin{equation}
 M_{\rm rec}=\sqrt{(\sqrt{s}-\sqrt{|\vec{p}_{D_s^-}|^2+m^2_{D_s^-}})^2-|\vec{p}_{D_s^-}|^2}, \nonumber
\end{equation}
where $m_{D^-_s}$ and $\vec{p}_{D_s^-}$ are the known mass~\cite{pdg24} and measured momentum of the tag $D_s^-$.
To suppress the other hadronic backgrounds from non-$D_s^*D_s$ processes, requirements on the kinematic variable $M_{\rm BC} = \sqrt{(\sqrt{s}/2)^2-|\vec {p}_{D_s^-}|^2}$ are applied. The detailed requirements on $M_{\rm BC} $ are described in Ref.~\cite{prd110_052012}; these requirements accept most ST candidates resulting from either the $D_s$ or $D_s^*$.  
To extract the mode-by-mode ST yields, we perform unbinned maximum likelihood fits to the distributions
of the $D^-_s$ invariant mass $M_{D_s^-}$, as shown in Fig.~\ref{fig:tag_mds}.
Signals are modeled with the MC-derived signal shape convolved with a Gaussian function to account for
the resolution differences between data and MC, while the
combinatorial backgrounds are parameterized with second-order polynomial
functions. Due to misidentification of $\pi^-$ as $K^-$,
the backgrounds from $D^-\rightarrow K^0_S\pi^-$ form a broad peak near the $D_s^-$ nominal mass for $D_s^-\rightarrow K^0_SK^-$.
In the fit, the shape of this background is described using the MC simulation, and the size of this background relative to other combinatorial ones is fixed. 
For each tag mode, the ST yield is obtained by integrating the signal
function over the $D^-_s$ signal region specified by $1.94<M_{D_s^-}<1.99$~GeV/$c^2$. The total ST yield summed over all fourteen ST modes is 
$N_{\rm ST}=(783.1\pm2.5)\times 10^3$, where the uncertainty is statistical only.

Candidates for the SL decay $D^+_s\rightarrow K^*(892)^0\ell^+\nu_{\ell}$ are selected from the remaining tracks recoiling against the ST $D^-_s$ mesons. The $K^*(892)^0$ meson is reconstructed via the $K^*(892)^0 \to K^+\pi^-$ decay, with the kaon and pion selected with the same criteria as applied on the ST side. It is required that there be exactly three charged tracks in the event in addition to the ST, and the invariant mass $M_{K^+\pi^-}$ is required to be within (0.801,0.991)~GeV/$c^2$. Detection and reconstruction of the positron follow the procedures in Ref.~\cite{prd110_052012}.
For muon identification, the specific energy loss $dE/dx$ and time-of-flight system (TOF) measurements are combined with shower properties from the electromagnetic calorimeter (EMC) to construct likelihoods for the electron ($\mathcal{L}_e$), muon ($\mathcal{L}_{\mu}$), and kaon ($\mathcal{L}_K$) hypotheses. The muon candidate must satisfy $\mathcal{L}_{\mu} > 0.001$, $\mathcal{L}_{\mu}>\mathcal{L}_e$, and $\mathcal{L}_{\mu}>\mathcal{L}_{K}$.
Additionally, the energy deposited in the EMC (${\rm EMC}_{\mu}$) of the muon candidate is required to be within $0.1<{\rm EMC}_{\mu}<0.3$~GeV. 
To suppress hadronic backgrounds, the invariant masses of $K^+\pi^-e^+$ and $K^+\pi^-\mu^+$ are required to be less than 1.80~GeV/$c^2$ and 1.60~GeV/$c^2$, respectively. 
Backgrounds containing additional $\pi^0$ mesons are further suppressed by requiring the maximum energy of any unused photon ($E_{\gamma \rm max}$) to be less than 0.20~GeV and 0.15~GeV for the $e$ and $\mu$ modes, respectively. 

To identify the photon produced directly from  the $D_s^{*\pm}$ decay, we perform two kinematic fits for each $\gamma$ candidate, one assuming that the $\gamma$ combines with the tag to form a $D_s^{*-}$ and the other assuming that the SL decay comes from a $D_s^{*+}$ parent. We require the $D^{\mp}_sD_s^{*\pm}$ pair to conserve energy and momentum in the center-of-mass frame, and the $D^{\pm}_s$ candidates are constrained to the known mass. The neutrino is treated as a missing particle. When we assume the tag to be the daughter of a $D_s^{*-}$, we constrain the mass of the photon plus tag candidate to be consistent with the known $D_s^{*-}$ mass; otherwise we constrain the mass of the photon plus SL decay to be consistent with the $D_s^{*+}$mass. Finally, we select the photon and hypothesis with the smallest kinematic fit $\chi^2_{\rm KF}$, this $\chi^2_{\rm KF}$ is also required be less than 50. 
We obtain information about the undetected neutrino with the missing-mass squared of an event, calculated by
the energies and momenta of the tag ($E_{D_s^-}$, $\vec{p}_{D_s^-}$), transition photon ($E_{\gamma}$,
$\vec{p}_{\gamma}$), and the detected SL products ($E_{\rm SL}=E_{K^*(892)^0}+E_{\ell^+}$,
$\vec{p}_{\rm SL}=\vec{p}_{K^*(892)^0}+\vec{p}_{\ell^+}$) by
\begin{equation}
{\rm MM}^2=(\sqrt{s}-E_{D_s^-}-E_{\gamma}-E_{\rm SL})^2-(|\vec{p}_{D_s^-}+\vec{p}_{\gamma}+\vec{p}_{\rm SL}|)^2. \nonumber
\end{equation}

\begin{figure}[tp!]
\begin{center}
   \includegraphics[width=\linewidth]{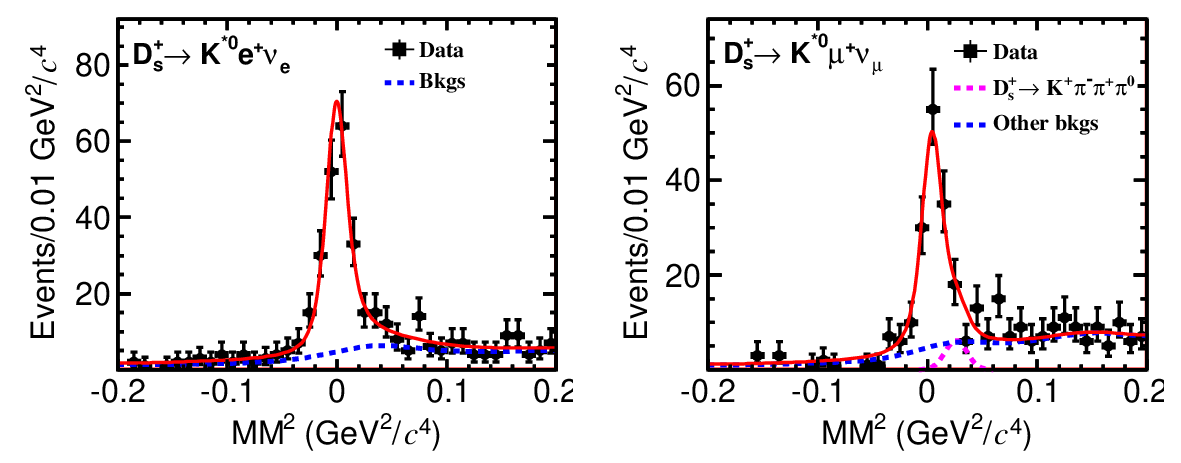}
   \caption{ (Color online)~Fits to ${\rm MM}^2$ distributions for (left) $D^+_s\rightarrow K^*(892)^0 e^+\nu_{e}$ and (right) $D^+_s\rightarrow K^*(892)^0 \mu^+\nu_{\mu}$.
   The dots with error bars are data, dot-dashed blue lines are the fitted combinatorial backgrounds and solid red curve are the total fits. The long-dashed pink line shows the fitted background from $D^+_s\rightarrow K^+\pi^-\pi^+\pi^0$. }
\label{fig:m2miss}
\end{center}
\end{figure}

Figure~\ref{fig:m2miss} shows the ${\rm MM}^2$ distributions of the accepted candidates for $D^+_s\rightarrow K^*(892)^0 e^+\nu_{e}$ and $D^+_s\rightarrow K^*(892)^0\mu^+\nu_{\mu}$ in data. The signal DT yields, $N_{\rm DT}$, are obtained by performing an unbinned maximum likelihood fits to the ${\rm MM}^2$ distributions.
In the fits, the signal is described with an MC-derived signal shape convolved with a Gaussian, 
and the combinatorial background is described by a shape obtained from the inclusive MC sample.
The residual peaking background from $D^+_s\rightarrow K^+\pi^-\pi^+\pi^0$ is simulated using the MC-derived shape, while its yield is fixed to be 15.5 based on the MC simulation.
We obtain $233.1\pm19.5$ and $133.6\pm14.1$ signal events for $D^+_s\rightarrow K^*(892)^0 e^+\nu_{e}$ and $D^+_s\rightarrow K^*(892)^0\mu^+\nu_{\mu}$, respectively. 
No peaking backgrounds are observed in either $K^{*0}$ and $D_{s}^{+}$ mass sidebands.
The average efficiencies $\epsilon_{\rm SL}$ of reconstructed $e$ and $\mu$ SL decays are estimated to be $(20.84\pm0.19)\%$ and $(12.39\pm0.12)\%$, respectively, where the BF of $K^*(892)^0\rightarrow K^+\pi^-$ is not included~\cite{BFK0}.

With the DT technique, the measured BFs of the $e$ and $\mu$ modes are insensitive to the systematic uncertainties of the ST selection.
The uncertainties of the $e^+$ and $\mu^+$ tracking and PID efficiencies are 1.0\% each, based on studies of $e^+e^-\rightarrow \gamma e^+e^-$ and $e^+e^-\rightarrow \gamma\mu^+\mu^-$ events, respectively, while the uncertainties due to the kaon and pion tracking and PID efficiencies are taken as 1.0\% each, using as control samples $D^0\rightarrow K^-\pi^+(\pi^0, \pi^+\pi^-)$ and $D^+\rightarrow K^-\pi^+\pi^+(\pi^0)$ decays~\cite{prl122_061801}. The uncertainty due to the $\gamma$ selection is estimated to be 1.0\% based on the requirement for the best photon candidate in a control sample of $e^+e^-\rightarrow D_s^{+*}D_s^-$ events with two hadronic decays, $D_s^+\rightarrow K_S^0K^+$ and $D_s^+\rightarrow K^+K^-\pi^+$.
The uncertainty associated with the ${\rm MM}^2$ fit for the $e~(\mu)$ decay is estimated to be 1.1~(1.4)\% by varying the fitting ranges, the signal and background shapes.
The systematic uncertainties due to the requirements of $E_{\rm \gamma max}$ and $\chi_{\rm KF}^{2}$
are estimated using the control samples $D_s^+\to K_S^0K^+$ and $D_s^+\to K_S^0K^+\pi^0$. 
The difference of the reconstruction efficiencies between data and MC due to the $E_{\rm \gamma max}$ requirement,
1.3~(2.0)\%, is assigned as the systematic uncertainty for the $e~(\mu)$ decay.
Similarly, a 1.9\% uncertainty is assigned due to the $\chi_{\rm KF}^{2}$ requirement for both decay modes.
The uncertainty due to the $M_{K^+\pi^-e^+}$ ($M_{K^+\pi^-\mu^+}$) requirement is estimated to be 0.8~(1.6)\% by comparing the nominal BF with that measured with alternative requirements.
The uncertainty due to the MC signal modeling is estimated to be 0.9\% by varying the input FF parameter by $\pm1\sigma$ measured in this work.
A 1.0\% systematic uncertainty from $N_{\rm ST}$ is evaluated using alternative signal shapes when fitting the $M_{D_s^-}$ spectra.  The uncertainty from the $K^*(892)^0$ mass window is estimated with $D^+\rightarrow K^*(892)^0\ell^+\nu_{\ell}$ decays is estimated as 0.3\%.  The MC statistics contribute an uncertainty of 0.9\%.
The uncertainty associated with the tag bias, introduced due to different reconstruction environments in the inclusive and signal MC samples, is estimated to be 0.2\%.
Adding these contributions in quadrature gives the total systematic uncertainty of 4.6\% and 5.1\% for the BFs of $e$ and $\mu$ modes of the $D_s^+$ SL decay, respectively. Finally, the BFs are determined to be $\mathcal B({D^+_s\rightarrow K^*(892)^0 e^+\nu_{e}})=(2.14\pm0.18_{\rm stat}\pm0.10_{\rm syst})\times10^{-3}$ and $\mathcal B({D^+_s\rightarrow K^*(892)^0 \mu^+\nu_{\mu}})=(2.07\pm0.22_{\rm stat}\pm0.10_{\rm syst})\times10^{-3}$.

The differential decay rate of $D^+_s\rightarrow K^*(892)^0 \ell^+\nu_{\ell}$ depends on five variables:
the $K^+\pi^-$ mass-squared ($M_{K^+\pi^-}^2$), the $\ell^+\nu_{\ell}$ mass-squared ($q^2$), the angle between the
$K^+$ and $D_s^+$ momenta in the $K^+\pi^-$ rest frame ($\theta_K$), 
the angle between the $\nu_{\ell}$ and $D^+_s$ momenta in the $\ell^+\nu_{\ell}$ system ($\theta_{\ell}$), and the acoplanarity angle between the $K^+\pi^-$ and $\ell^+\nu_{\ell}$ decay planes ($\chi$).
The differential decay rate can be expressed in terms of three helicity amplitudes~\cite{RevModPhys67_893,prd46_5040},
\begin{eqnarray}
  H_{\pm}(q^2) &=& (M_{D^+_s}+m_{K\pi})A_1(q^2) \nonumber \\
   & &\mp\frac{2p_{K\pi}M_{D^+_s} }{M_{K\pi}+M_{D^+_s}} \, V(q^2) \nonumber \\
  H_0(q^2) &=& \frac{1}{2qm_{K\pi}} \nonumber \\
           & & \left[(M_{D^+_s}^2-m_{K\pi}^2-q^2)(m_{K\pi}+M_{D^+_s})A_1(q^2)  \right. \nonumber \\
           & & \left.-\frac{4p^2_{K\pi}M_{D^+_s}^2}{M_{K\pi}+M_{D^+_s}} \, A_2(q^2)\right] \, , \nonumber
\end{eqnarray}
where $p_{K\pi}$ is the momentum of the $K^+\pi^-$ system in the rest frame of the $D_s^+$,
and $V(q^2)$ and $A_{1,2}(q^2)$ are the vector and axial FFs, respectively.
Since $A_1(q^2)$ is common to all three helicity amplitudes, this FF is used as a normalization in the FF ratios $r_V=V(0)/A_1(0)$ and $r_2=A_2(0)/A_1(0)$.
The FFs $A_{1,2}(q^2)$ and $V(q^2)$ are assumed to have simple pole forms,
$A_{1,2}(q^2)=A_{1,2}(0)/(1-q^2/M^2_A)$ and $V(q^2)=V(0)/(1-q^2/M^2_V)$, with pole masses
$M_V=M_{D^*(2010)}=2.01$~GeV/$c^2$ and $M_A=M_{D_1(2420)}=2.42$~GeV/$c^2$~\cite{pdg24}.

\begin{figure}[tp!]
\begin{center}
   \includegraphics[width=\linewidth]{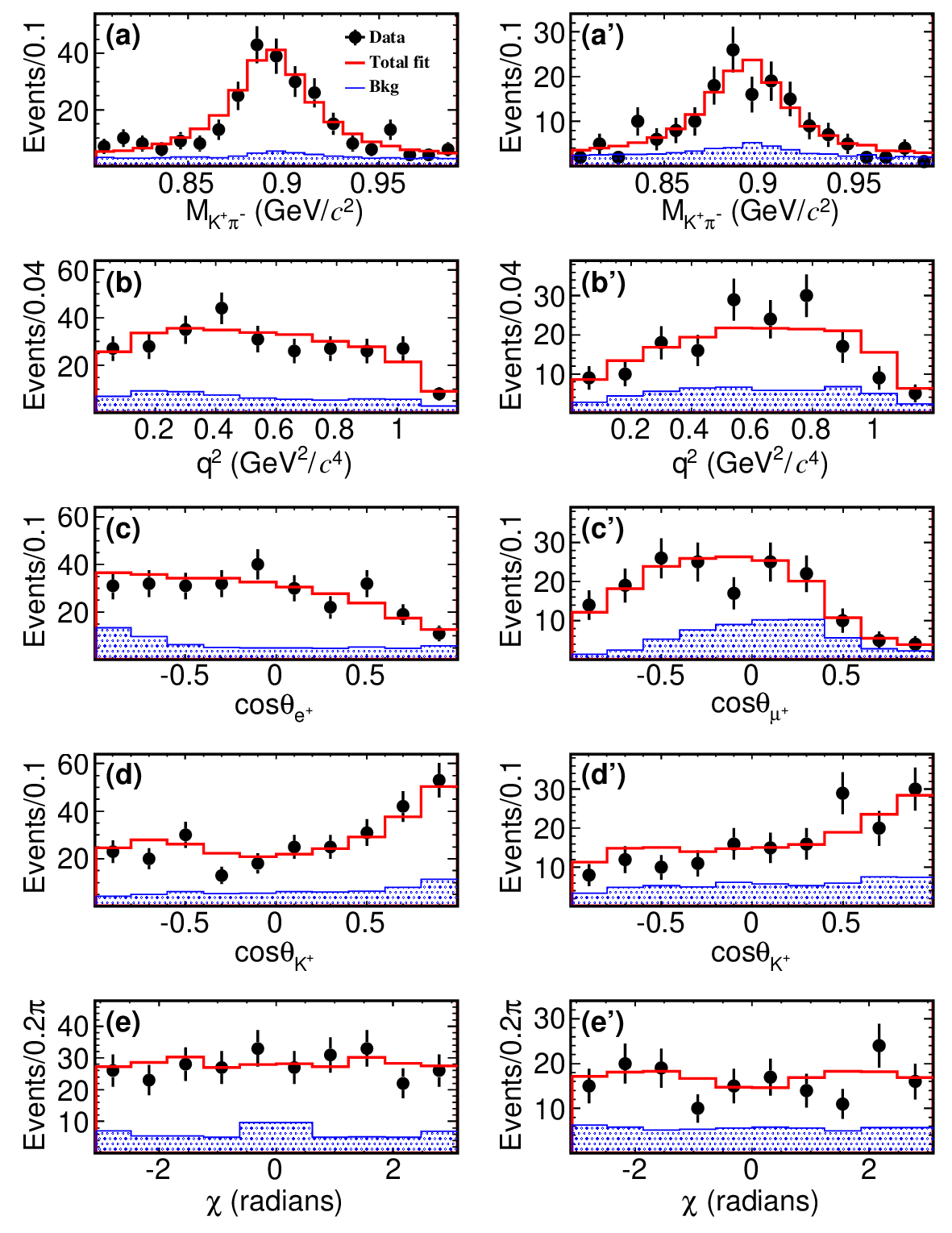}
   \caption{ (Color online)~ Projections of (a, a$^{\prime}$) $M_{K^+\pi^-}$, (b, b$^{\prime}$)  $q^2$, (c, c$^{\prime}$) $\cos\theta_{\ell^+}$, (d, d$^{\prime}$) $\cos\theta_{K^+}$, and (e, e$^{\prime}$)  $\chi$ for (left) $D^+_s\rightarrow K^*(892)^0 e^+\nu_{e}$ and (right) $D^+_s\rightarrow K^*(892)^0 \mu^+\nu_{\mu}$. The dots with error bars are data, red histograms are the fit results, and shaded histograms are the simulated background. }
\label{fig:formfactor}
\end{center}
\end{figure}

\begin{figure}[htbp]
   \begin{center}
   \includegraphics[width=\linewidth]{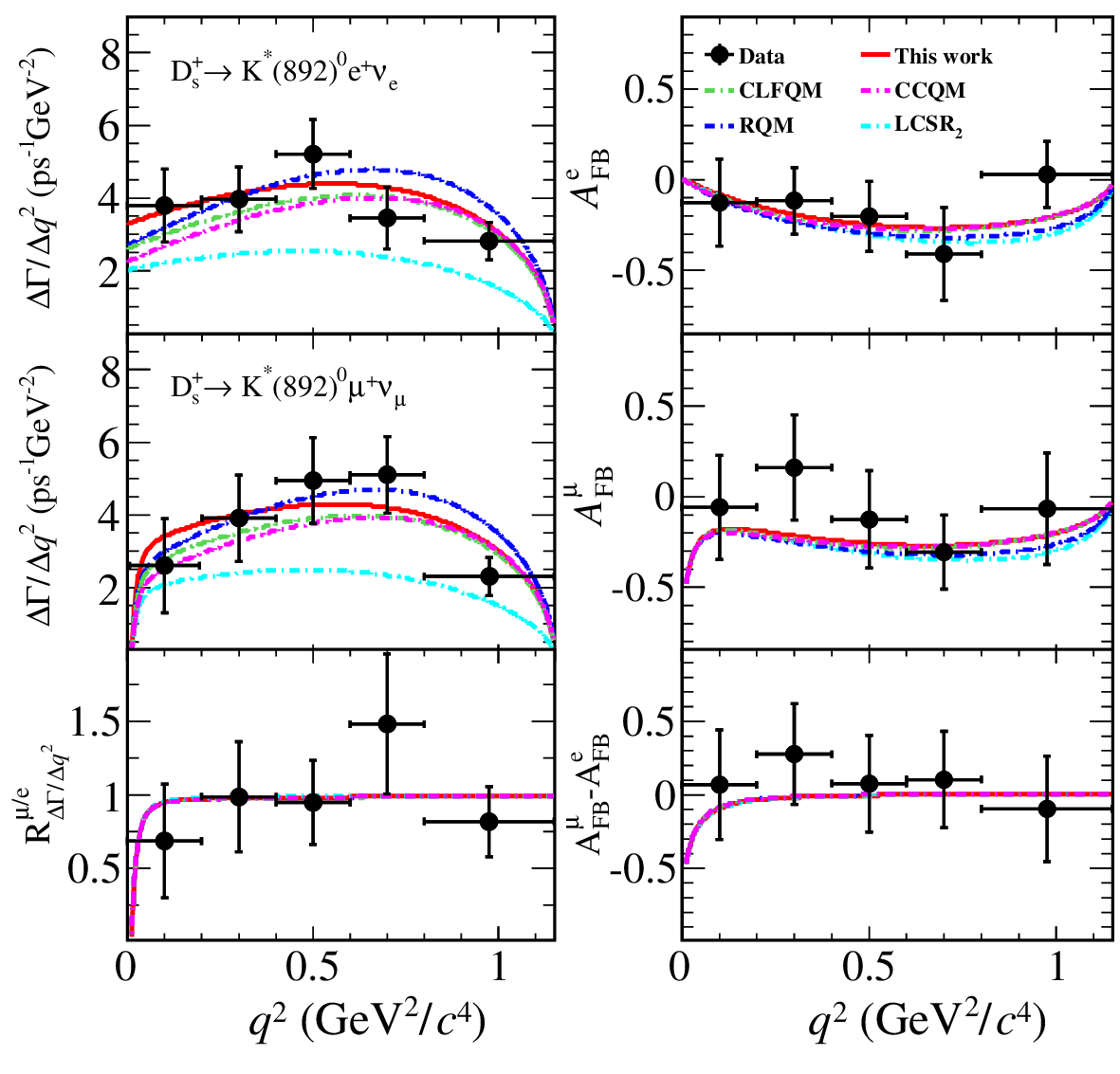}
   \caption{Top Row: Measurements of (left) $\Delta\Gamma/\Delta q^2$ and (right) $A^{\ell}_{\rm FB}$ vs.~$q^2$ for $D_s^+\rightarrow K^*(892)^0 e^+\nu_e$.  
Middle Row: the same quantities for  $D_s^+\rightarrow K^*(892)^0 \mu^+\nu_{\mu}$. Bottom Row: the ratio $\mathcal{R}^{\mu/e}_{\Delta\Gamma/\Delta q^2}$ and the difference $A^{\mu}_{\rm FB}-A^{e}_{\rm FB}$. The dots with error bars are data, where statistical and systematic uncertainties are both included. The red curves show the derived central values using FFs measured in this work, while the green, pink, blue and cyan curves show the values calculated with FF values predicted by the CLFQM~\cite{JHEP02_179}, the CCQM~\cite{PRD98_114031}, the RQM~\cite{PRD101_013004}, and the LCSR$_{2}$~\cite{PRD111_113005}, respectively. }
   \label{fig:Decayrates}
\end{center}
\end{figure}

To determine the FF parameters, a simultaneous five-dimensional maximum likelihood fit is performed in the space of $(M^2_{K^+\pi^-}, q^2,
\cos\theta_{\ell}, \cos\theta_{K^+}, \chi)$ for the $D^+_s\rightarrow K^*(892)^0e^+\nu_{e}$ and $D^+_s\rightarrow K^*(892)^0\mu^+\nu_{\mu}$ decays.
A similar approach is used in Refs.~\cite{prl110_131802,prd92_071101} where the events within $-0.08<{\rm MM}^2<0.08$~GeV$^2$/$c^4$ and $-0.04<{\rm MM}^2<0.04$~GeV$^2$/$c^4$ are selected for $e$ and $\mu$ modes, respectively.
The significance of the $\mathcal{S}$-wave component in the $K^+\pi^-$ system is estimated to be less than $2\sigma$; it is neglected in the fit.
The $K^*(892)^0$ component is described with a relativistic Breit-Wigner function following Ref.~\cite{D0Kspiev}, with the mass and width determined as free fit parameters.
The projections for the fit onto $M^2_{K^+\pi^-}$, $q^2$, $\cos\theta_{\ell}$, $\cos\theta_{K^+}$, and $\chi$ are shown in
Fig.~\ref{fig:formfactor}. The fit result is $r_V=1.63\pm0.14_{\rm stat}$ and $r_2=0.60\pm0.13_{\rm stat}$.
The $K^*(892)^0$ mass and width are fitted to be $895.0\pm1.7$~MeV/$c^2$ and $46.3\pm4.1$~MeV, respectively, both consistent with PDG averages~\cite{pdg24}.
The fit procedure has been validated by analyzing a large inclusive MC sample.
The systematic uncertainties in the measurement of FF ratios are estimated by comparing the nominal values with
those obtained after varying each source of uncertainty, as described in Ref.~\cite{D0Kspiev}.
The systematic uncertainties for $r_V$\,($r_2$) mainly arise from the uncertainties related to the tracking, PID and photon detection [1.7\,(4.2)\%], the $K^{*0}$ mass window [1.4\,(3.8)\%],
the $E_{\gamma{\rm max}}$ requirement [1.8\,(4.2)\%], the
$M_{K^+\pi^-e^+/\mu^+}$ requirement [3.0\,(5.5)\%], the $\chi_{\rm KF}^2$ requirement [2.3\,(2.2)\%], background estimation [0.8\,(1.4)\%], and
the possible $\mathcal{S}$-wave contribution [0.7\,(1.0)\%].
Combining all of these in quadrature, the systematic uncertainties for the $r_V$ and $r_2$ of $D_s^+\rightarrow K^*(892)^0\ell^+\nu_{\ell}$ are estimated to be 4.9\% and 9.5\%, respectively. Therefore, the measured value of FF parameters are $r_V=1.63\pm0.14_{\rm stat}\pm0.08_{\rm syst}$ and $r_2=0.60\pm0.13_{\rm stat}\pm0.06_{\rm syst}$.

\begin{figure*}[tp!]
\begin{center}
   \begin{minipage}[t]{5.8cm}
   \includegraphics[width=\linewidth]{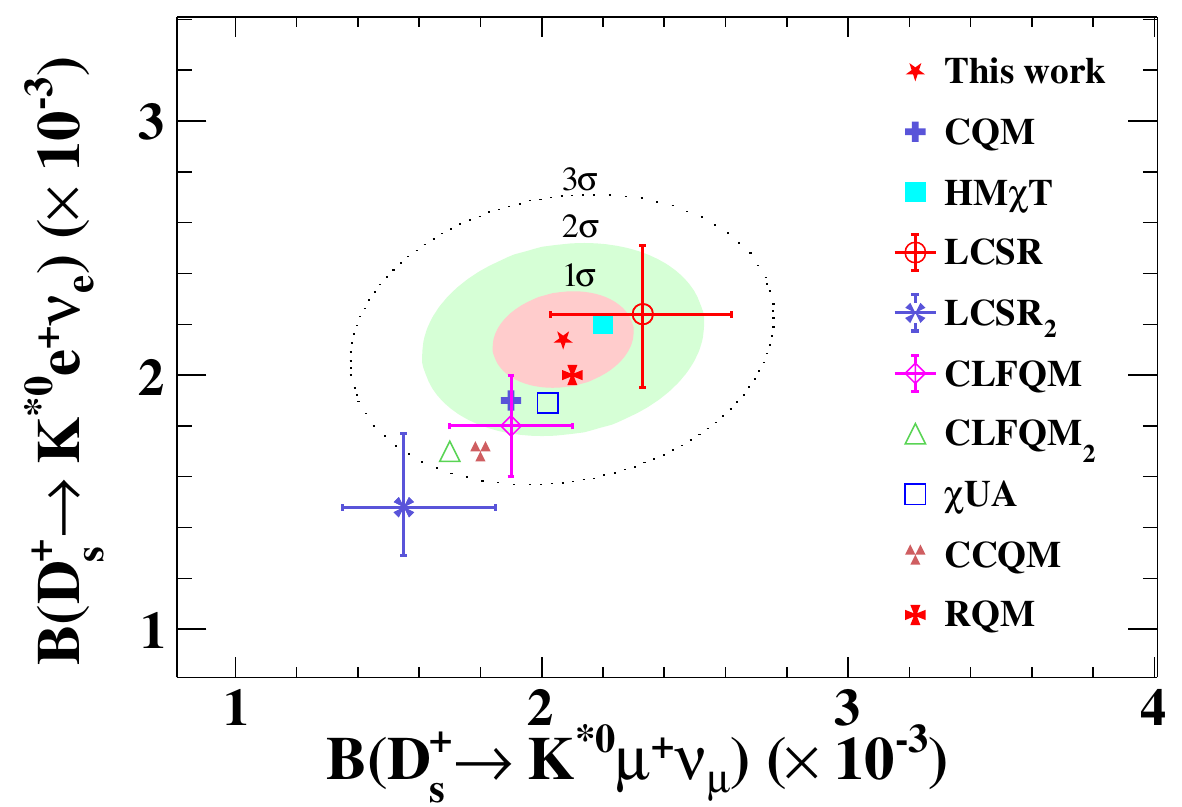}
   \end{minipage}    
   \begin{minipage}[t]{5.8cm}
   \includegraphics[width=\linewidth]{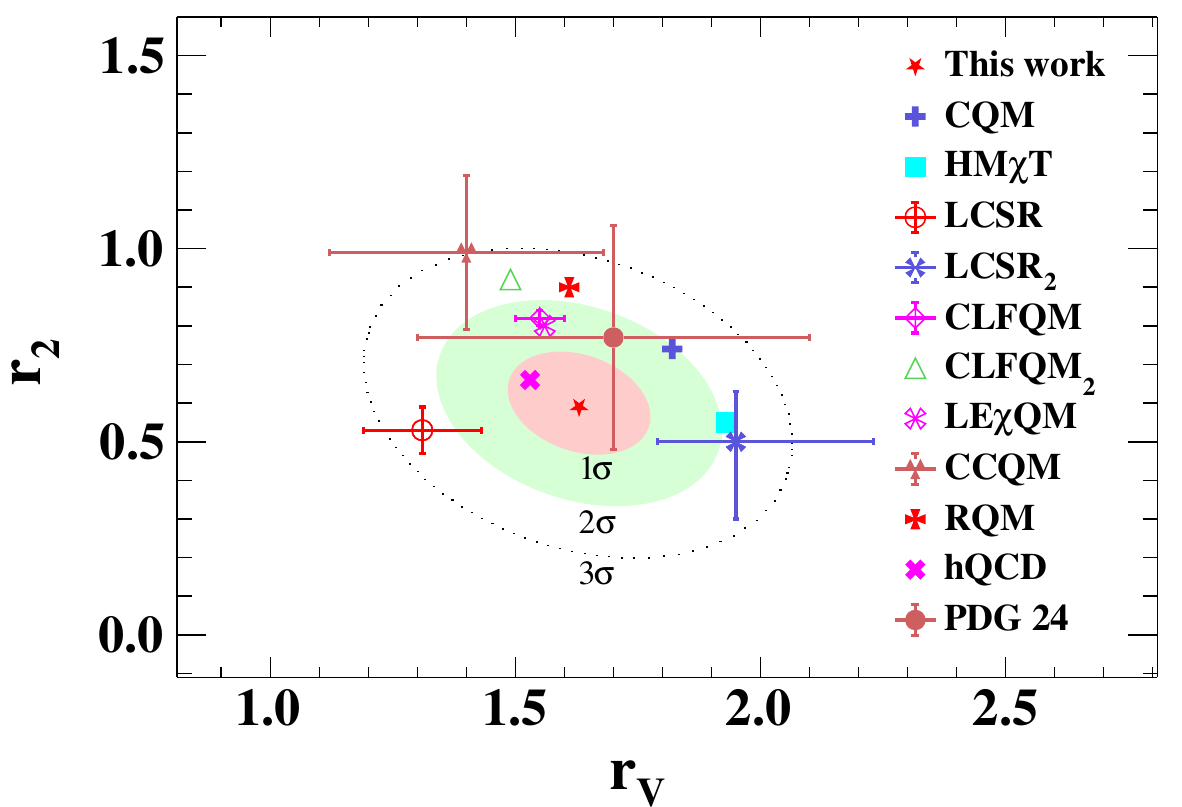}
   \end{minipage} 
   \begin{minipage}[t]{5.8cm}
   \includegraphics[width=\linewidth]{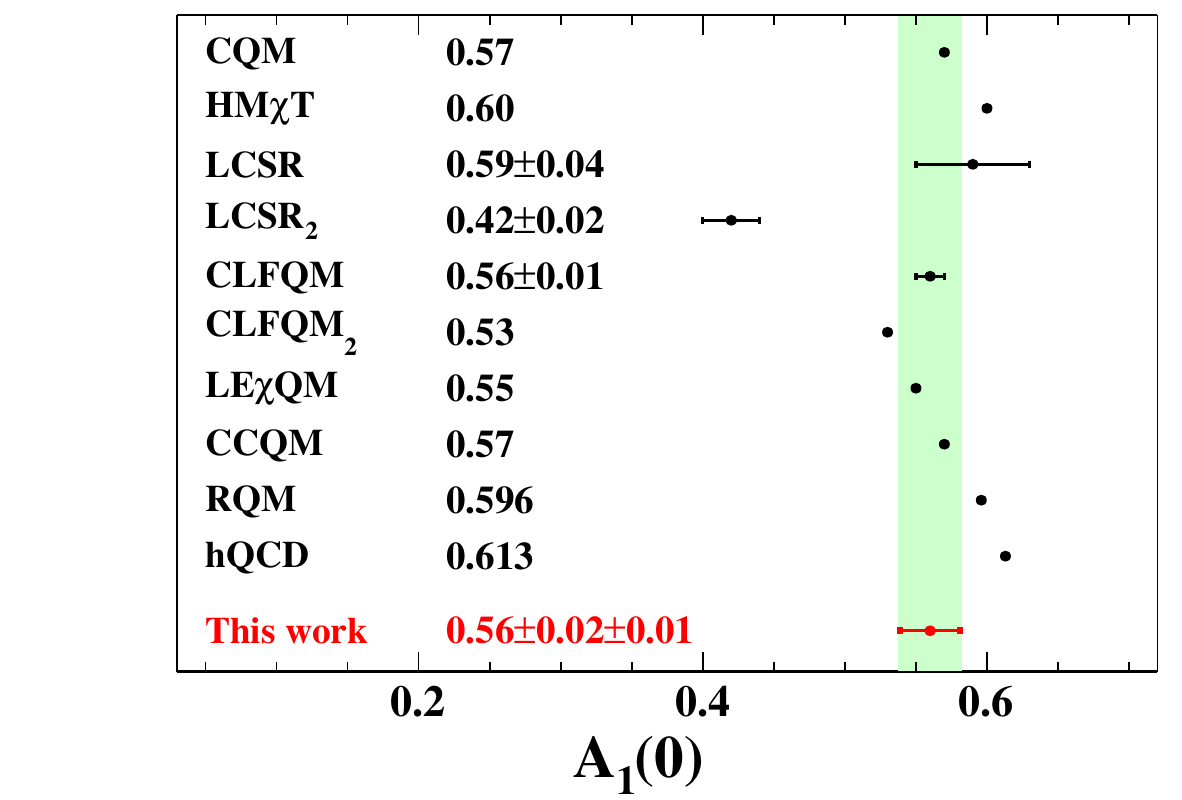}
   \end{minipage}                   
   \caption{(Color online)~Comparisons of the measured (left) BFs, (middle) $r_2$ and $r_V$, and (right) $A_1(0)$ for the $D^+_s\rightarrow K^*(892)^0 e^+\nu_{e}$ and $D^+_s\rightarrow K^*(892)^0 \mu^+\nu_{\mu}$ decays, with theoretical calculations from the CQM~\cite{PRD62_014006}, HM$\chi$T~\cite{PRD72_034029}, LCSR~\cite{IJMPA21_6125}, LCSR$_{2}$~\cite{PRD111_113005}, CLFQM~\cite{JPG39_025005,EPJC77_587,JHEP02_179}, CLFQM$_{2}$~\cite{PRD78_054002}, LE$\chi$QM~\cite{PRD89_034013}, $\chi$UA~\cite{PRD92_054038}, CCQM~\cite{PRD98_114031,FrontPhys14_64401}, RQM~\cite{PRD101_013004}, hQCD~\cite{PRD109_026008}, and the PDG 2024 experimental average~\cite{pdg24}. The correlation coefficients between the measured BFs and FFs are $0.15$ and $-0.28$, respectively. }
\label{fig:cmpbf}
\end{center}
\end{figure*}

We also perform the first model-independent measurements for the partial decay rate $\Delta\Gamma_i/\Delta q^2$, and the forward-backward asymmetries $A^{\ell}_{\rm FB}(q^2)$ for the decays
$D^+_s\rightarrow K^*(892)^0 \ell^+\nu_{\ell}$. The observables are measured by partitioning the signal candidates into five $q^2$ intervals defined as [0,~0.20), [0.20,~0.40), [0.40,~0.60), [0.60,~0.80), [0.80,~$q^2_{\rm max}$)~GeV$^2$/$c^4$, where $q^2_{\rm max} = (M_{D_s^+}-M_{K^*(892)^0})^2$, and $M_{D_s^+}$ and $M_{K^*(892)^0}$ are the known masses of the $D_s^+$ and $K^*(892)^0)$~\cite{pdg24}. The partial decay rate in the $i^{th}$ $q^2$ interval, $\Delta \Gamma_i$, is determined as
\begin{equation}
\Delta \Gamma_i=\int_i \frac{d\Gamma}{dq^2} \, dq^2 = \sum\limits_{j=1}^{N_{\rm bins}}(\epsilon^{-1})_{ij}N^j_{\rm DT}/(\tau_{D_s^+} \, N^{\rm ST}),
\label{eq:drate}
\end{equation}
where $\tau_{D_s^+}$ is the lifetime of $D^+_s$~\cite{pdg24}, and $\epsilon_{ij}$ is the efficiency matrix describing the reconstruction efficiency of SL decays including migration between $q^2$ bins~\cite{PRL124_241802,PRD101_112002}. 
The $A^{\ell}_{\rm FB}(q^2)$ are measured in the same manner to Ref.~\cite{PRD108_L031105}
\begin{equation}
A^{\ell}_{\rm FB}(q^2) = \frac{\int_{0}^{1}\frac{d^2\Gamma}{dq^2 d \cos\theta_{\ell}} \, d\cos\theta_{\ell}-\int_{-1}^{0}\frac{d^2\Gamma}{dq^2d\cos\theta_{\ell}} \, d\cos\theta_{\ell}}{\int^{1}_{0}\frac{d^2\Gamma}{dq^2d\cos\theta_{\ell}} \, d\cos\theta_{\ell}+\int_{-1}^{0}\frac{d^2\Gamma}{dq^2d\cos\theta_{\ell}} \, d\cos\theta_{\ell}},
\label{eq:AFB}
\end{equation}
where $d^2\Gamma/(dq^2 \, d\cos\theta_{\ell})$ is the two-dimensional differential decay rate. 
To measure $\Delta\Gamma_i/\Delta q^2$ and $A^{\ell}_{\rm FB}(q^2)$, the signal yield in the $j^{th}$ $q^2$ interval, \emph{i.e.} the yields $N^j_{\rm DT}$ are
obtained from fits to the corresponding ${\rm MM}^2$ distribution. Taking into account the correction due to detection efficiency (Eq.~\ref{eq:drate}), the differential decay rates, $\Delta\Gamma_i/\Delta q^2$, for the decays $D^+_s\rightarrow K^*(892)^0 e^+\nu_e$ and $D^+_s\rightarrow K^*(892)^0 \mu^+\nu_{\mu}$ are measured. We also measure $A^{\ell}_{\rm FB}(q^2)$ using $\Delta\Gamma_i/\Delta q^2$ extracted within the regions $\cos\theta_{\ell}\in(0,~1)$ and $\cos\theta_{\ell}\in(-1,~0)$, respectively.
The results of $\Delta\Gamma_i/\Delta q^2$ and $A^{\ell}_{\rm FB}(q^2)$ for $D^+_s$ SL decays as well as the ratio $\mathcal{R}^{\mu/e}_{\Delta \Gamma/\Delta q^2}$ and the difference $A^{\mu}_{\rm FB}-A^{e}_{\rm FB}$ between $\mu$ and $e$ modes, are shown in Fig.~\ref{fig:Decayrates}.
Comparisons of our measurements with theoretical calculations in separate $q^2$ intervals show no violation of the LFU within experimental uncertainties.

This Letter reports the first complete measurement of the dynamics in the $D^+_s \rightarrow K^*(892)^0\ell^+\nu_{\ell}$ decays.
The BFs of $e$ and $\mu$ modes as well as their ratio are measured to be $\mathcal B({D^+_s\rightarrow K^*(892)^0 \mu^+\nu_{\mu}})=(2.07\pm0.22_{\rm stat}\pm0.10_{\rm syst})\times10^{-3}$,
$\mathcal B({D^+_s\rightarrow K^*(892)^0 e^+\nu_{e}})=(2.14\pm0.18_{\rm stat}\pm0.10_{\rm syst})\times10^{-3}$, and $R^{\mu/e}_{K^*(892)}=0.97\pm0.13_{\rm stat}\pm0.04_{\rm syst}$, respectively.
Based on a simultaneous analysis of the dynamics in the $D^+_s \rightarrow K^*(892)^0\ell^+\nu_{\ell}$ decays, we report the most precise determination of the FF parameters in the $D_s^+\rightarrow K^*(892)^0$ transition and measure $r_V=1.63\pm0.14_{\rm stat}\pm0.08_{\rm syst}$ and $r_2=0.60\pm0.13_{\rm stat}\pm0.06_{\rm syst}$ at $q^2=0$. 
Using $|V_{cd}|=0.22487\pm0.00068$~\cite{pdg24} and $\tau_{D_s^+}=(501.2\pm2.2)$~fs~\cite{pdg24}, we obtain $A_1(0)=0.56\pm0.02_{\rm stat}\pm0.01_{\rm syst}$, which represents the first determination for the $D_s^+\rightarrow K^*(892)^0$ transition. 
The measured BFs and FFs presented in this work provide stringent tests on various theoretical models as shown on Fig.~\ref{fig:cmpbf}. At a confidence level  (C.L.) of 95\%, our measured BFs disfavor the central values calculated with the LCSR$_{2}$~\cite{PRD111_113005}, CLFQM$_{2}$~\cite{PRD78_054002}, and CCQM~\cite{PRD98_114031,FrontPhys14_64401}. The measured FF parameters $r_V$ and $r_2$ disfavor the central value calculated with the CLFQM$_{2}$~\cite{PRD78_054002}, and the measured $A_1(0)$ disfavors the central values calculated with the LCSR$_{2}$~\cite{PRD111_113005} and the hQCD~\cite{PRD109_026008}.
The significant deviation of the LCSR calculation in Ref.~\cite{PRD111_113005} with respect to our measured $A_1(0)$ suggests big theoretical challenges when theory incorporates contributions of meson distribution amplitudes with higher-twist corrections. Our results in this Letter provide an urgent calibration in data.

\begin{figure}[tp!]
\begin{center}
   \begin{minipage}[t]{5.5cm}
   \includegraphics[width=\linewidth]{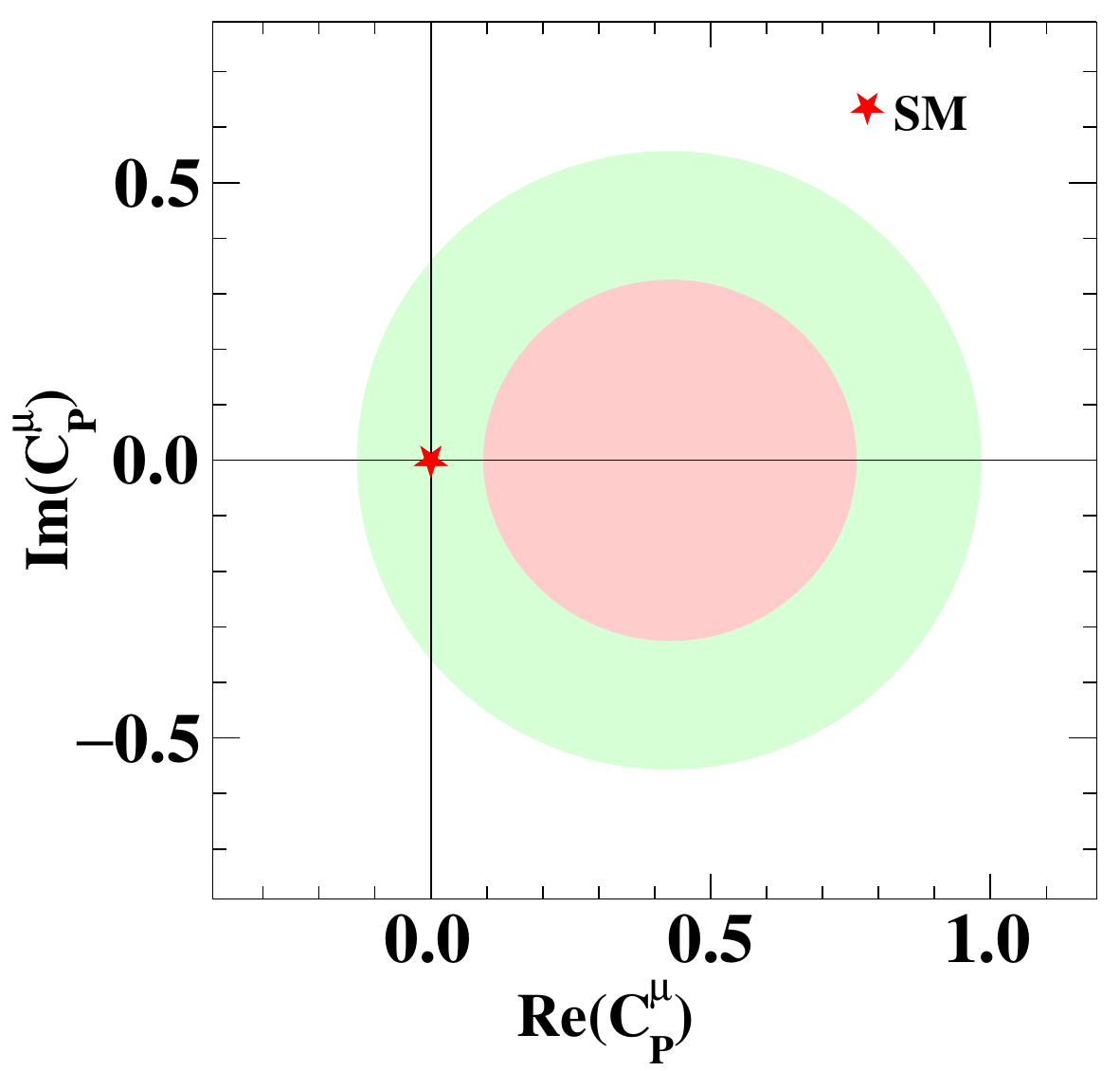}
   \end{minipage}                        
   \caption{(Color online)~Allowed region of extracted $C_P^{\mu}$. The 68\% and 95\% C.L. regions are shown in red and green shades. }
\label{fig:cplemu}
\end{center}
\end{figure}

With the FF parameters measured in this work, we also determine the ratios of partial widths for longitudinal and transverse $K^*(892)^0$ polarizations~\cite{PRD91_094009,EPJC77_587} for $e$ and $\mu$ modes to be $R^{e}_{L/T} = [\Gamma_L/\Gamma_T]^e = 1.31 \pm 0.10_{\rm stat} \pm 0.03_{\rm syst}$ and $R^{\mu}_{L/T} = [\Gamma_L/\Gamma_T]^{\mu} = 1.25 \pm 0.09_{\rm stat} \pm 0.03_{\rm syst}$, respectively. 
Combining the measured $R^{\mu}_{L/T}$ in this work and $R^{\mu}_{L/T}=1.21 \pm 0.05$ from the $D^0\rightarrow K^*(892)^-\mu^+\nu_{\mu}$ decay~\cite{PRL135_111803}, 
the pseudoscalar Wilson coefficient $C_P^{\mu}$ can be extracted as in Ref.~\cite{PRD91_094009}. The allowed region of $C_P^{\mu}$ is shown in Fig.~\ref{fig:cplemu}. This is obtained for the first time in $D_{(s)}\rightarrow K^*(892)\mu^+\nu_{\mu}$ decays and provides additional data to constrain the pseudoscalar new physics coefficient in the charged current $c\rightarrow (s,d)\bar{\ell}\nu_{\ell}$ transitions~\cite{EPJC80_153}. Using $r^{D\rightarrow \rho}_V=1.53\pm0.08$ and $r^{D\rightarrow \rho}_2=0.82\pm0.05$ measured in $D\rightarrow \rho \ell^+\nu_{\ell}$ decays~\cite{pdg24}, we determine the FF ratios of $D^+_s \rightarrow K^*(892)^0\ell^+\nu_{\ell}$ to $D\rightarrow \rho \ell^+\nu_{\ell}$ to be $r_V^{D^+_s \rightarrow K^*(892)^0}/r_V^{D\rightarrow \rho}=1.06\pm0.12$ and $r_2^{D^+_s \rightarrow K^*(892)^0}/r_2^{D\rightarrow \rho}=0.73\pm0.18$, respectively.
This $r_V$ ($r_2$) ratio is consistent within $1\sigma$ ($1.5\sigma$) with the LQCD prediction~\cite{lattice} and the expectation of $U$-spin ($d\leftrightarrow s$) symmetry~\cite{plb492_297}.
Our measurements provide a unique test of the LQCD prediction that the FFs are insensitive to spectator quarks, which has important implications when considering the corresponding SL decays of $B$ and $B_s$ mesons~\cite{lattice,EPJC74_2981,prd85_114502}.

We also provide the first model-independent measurement of the differential decay rates and the forward-backward asymmetries in both the full and binned four-momentum transfer regions for $D^+_s \rightarrow K^*(892)^0\ell^+\nu_{\ell}$ decays. Comparisons with various non-perturbative theoretical calculations are also shown in Fig.~\ref{fig:Decayrates}.
The detailed examination of their ratios show no evidence for LFU violation. Averaging the $A^{e}_{\rm FB}$ and $A^{\mu}_{\rm FB}$ measured in each $q^2$ intervals,
we determine the mean values of forward-backward asymmetries for $e$ and $\mu$ modes to be $\left\langle A^{e}_{\rm FB}\right\rangle=-0.14\pm0.09_{\rm stat}\pm0.01_{\rm syst}$ and $\left\langle A^{\mu}_{\rm FB}\right\rangle=-0.12\pm0.10_{\rm stat}\pm0.01_{\rm syst}$. 
These results are consistent within $1\sigma$ with the predictions of $-0.22$ to $-0.26$~\cite{FrontPhys14_64401,PRD101_013004,PRD98_114031,JHEP02_179} for the positron mode and $-0.25$ to $-0.29$~\cite{FrontPhys14_64401,PRD101_013004,PRD98_114031,JHEP02_179} for the muon mode, respectively. 
The results presented in this Letter provide powerful tests and constraints on various theoretical calculations, especially in QCD theories, and play an important role in understanding the dynamics of SL decays of charmed hadrons in the non-perturbative region.

\acknowledgments

The BESIII Collaboration thanks the staff of BEPCII (https://cstr.cn/31109.02.BEPC) and the IHEP computing center for their strong support. This work is supported in part by National Key R\&D Program of China under Contracts Nos. 2023YFA1606000, 2023YFA1606704; National Natural Science Foundation of China (NSFC) under Contracts Nos. 11635010, 11935015, 11935016, 11935018, 12025502, 12035009, 12035013, 12061131003, 12192260, 12192261, 12192262, 12192263, 12192264, 12192265, 12221005, 12225509, 12235017, 12342502, 12361141819, 12375090; the Chinese Academy of Sciences (CAS) Large-Scale Scientific Facility Program; the Strategic Priority Research Program of Chinese Academy of Sciences under Contract No. XDA0480600; CAS under Contract No. YSBR-101; 100 Talents Program of CAS; The Institute of Nuclear and Particle Physics (INPAC) and Shanghai Key Laboratory for Particle Physics and Cosmology; ERC under Contract No. 758462; German Research Foundation DFG under Contract No. FOR5327; Istituto Nazionale di Fisica Nucleare, Italy; Knut and Alice Wallenberg Foundation under Contracts Nos. 2021.0174, 2021.0299, 2023.0315; Ministry of Development of Turkey under Contract No. DPT2006K-120470; National Research Foundation of Korea under Contract No. NRF-2022R1A2C1092335; National Science and Technology fund of Mongolia; Polish National Science Centre under Contract No. 2024/53/B/ST2/00975; STFC (United Kingdom); Swedish Research Council under Contract No. 2019.04595; U. S. Department of Energy under Contract No. DE-FG02-05ER41374. This paper is also supported by the Fundamental Research Funds for the Central Universities, and the Research Funds of Renmin University of China under Contract No. 24XNKJ05.



\begin{thebibliography}{99}
\bibitem{BFK0} In this work, the BF of $K^*(892)^0\rightarrow K^+\pi^-$ is taken as 2/3 according to the hypothesis of isospin symmetry.
\end{thebibliography}
\end{document}